\documentclass[11pt,a4paper]{article}

\usepackage{graphicx} 
\usepackage{subcaption}
\usepackage{xcolor}
\usepackage{amsmath} 
\usepackage{amssymb} 
\usepackage{cite}
\usepackage{float}
\usepackage{braket}
\usepackage[T1]{fontenc}
\usepackage{titlesec}
\titleformat{\paragraph}
{\normalfont\normalsize\bfseries}{\theparagraph}{1em}{}
\titlespacing*{\paragraph}
{0pt}{3.25ex plus 1ex minus .2ex}{1.5ex plus .2ex}
\newcommand{\be}{\begin{equation}}
\newcommand{\ee}{\end{equation}}
\newcommand{\bpm}{\begin{pmatrix}}
\newcommand{\epm}{\end{pmatrix}}

\newcommand{\dd}{\mathrm{d}}
\newcommand{\tE}{\Tilde{E}}
\newcommand{\dvol}{\mathrm{d}\widetilde{\mathrm{vol}}}

\usepackage{tabularx} 
\usepackage{psfrag} 

\usepackage{eurosym} 
\def\?{\euro{}}
\usepackage{footnote}
\usepackage{tablefootnote}

\usepackage{color} 
\usepackage{mathtools}

\definecolor{linkcolor}{rgb}{0,0,0.6} 
\usepackage[pdftex,colorlinks=true,
pdfstartview=FitV,
linkcolor= linkcolor,
citecolor= linkcolor,
urlcolor= linkcolor,
hyperindex=true,
hyperfigures=false]
{hyperref} 

\usepackage{fancyhdr} 

\numberwithin{equation}{section}
\begin{document}

\begin{titlepage}

	\begin{center}

	\vskip .5in 
	\noindent

	{\Large \bf{A weakly non-abelian decay channel}}

	\bigskip\medskip

	 Vincent Menet and Alessandro Tomasiello \\

	\bigskip\medskip
	{\small 
Dipartimento di Matematica, Universit\`a di Milano--Bicocca, \\ Via Cozzi 55, 20126 Milano, Italy \\ and \\ INFN, sezione di Milano--Bicocca
		}

	\vskip .5cm 
	{\small \tt vincent.menet, alessandro.tomasiello@unimib.it}
	\vskip .9cm 
	     	{\bf Abstract }
	\vskip .1in
	\end{center}

We investigate non-abelian branes in curved space. 
We discuss solutions to the equations of motion of the transverse scalars when they are constant along the world-volume directions and obey an $\mathfrak{su}(2)$ or an $\mathfrak{su}(2)\oplus\mathfrak{su}(2)$ algebra.
Motivated by the membrane version of the weak gravity conjecture, we specialise our results to non-abelian domain-wall D$(d-2)$ branes embedded in AdS$_d$ flux vacua. We find that they can be less self-attractive than their abelian counterpart, opening up a new decay-channel for vacua that resist all abelian domain-wall destabilisations. These branes come in two types,
depending on whether their fuzziness involves the radial direction, or is purely internal.
Only the latter can develop in vacua free from abelian decays. We illustrate our construction by embedding these branes in a variety of AdS vacua, destabilising some of them.

	\vfill
	\eject

	\end{titlepage}

\tableofcontents

\newpage

\section{Introduction}


Most AdS vacua in string theory appear to suffer from instabilities. The lower-codimension version of the weak gravity conjecture \cite{Arkani-Hamed:2006emk} would indeed suggest  \cite{Ooguri:2016pdq}, if valid, that such vacua must admit a charged brane with a tension smaller than its
charge, which would nucleate and trigger vacuum decay via its expansion.

The investigation of the non-perturbative stability of AdS vacua has recently flourished \cite{Danielsson:2017max,Bena:2020xxb,Guarino2020BranejetSO,Suh:2020rma,Suh:2021icf,Bomans:2021ara,Marchesano:2021ycx,Marchesano:2022rpr,Menet:2025nbf}. From a string theory perspective, these domain-wall charged branes could correspond to a variety of objects. However, until now the literature has mainly focused on abelian domain-wall D-branes and their bound states. In particular, in our previous paper \cite{Menet:2025nbf} we found AdS solutions that are not destabilized by any abelian branes.

In this paper, we consider the scenario where this charged brane is actually a stack of $N$ coincident non-abelian domain-wall D$p$-branes. This results in the promotion of the scalars associated to the directions transverse to the branes to non-commuting matrices living in the adjoint representation of the world-volume gauge group $U(N)$. This limits the possible algebras obeyed by the scalars to reductive Lie ones, which only allows for three cases: $\mathfrak{su}(2)\oplus\mathbb{R}^{6-p},$ $\mathfrak{su}(2)\oplus\mathfrak{su}(2)\oplus\mathbb{R}^{3-p},$ and $\mathfrak{su}(3)\oplus\mathbb{R}^{1-p}$ if $p\leq6,3,1$ respectively.

A proposal for the DBI and CS actions for such branes has first been written down in \cite{Myers:1999ps}. Crucially, this action matches the open-string amplitude results---and is therefore reliable---only up to order $\alpha'^2$ \cite{Bergshoeff:2001dc,Bilal:2001hb}. Beyond this order, no closed form expression is currently known for the brane actions. The Myers action has been used to unravel the celebrated Myers effect, which highlights the ``polarisation'' of D$0$-branes in flat space, puffing up into D$2$-branes in the presence of four-form RR fluxes. The resulting non-abelian configuration obeys an $\mathfrak{su}(2)\oplus\mathbb{R}^{6}$ algebra, which is why its geometry is referred to as the \emph{fuzzy sphere}.

Even though our motivations originate from domain-wall D-branes in AdS, we first perform the completely general expansion of the Myers action in curved space, since it has been scarcely discussed. We carry this out in the \textit{weakly} non-abelian regime, by which we mean that we truncate the non-abelian contributions to the action to the order $\alpha'^2$.  These general results might have wider applicability for other issues regarding non-abelian bound states.

For the applications in this paper, we focus on the case of branes with scalars that are constant along the branes world-volume. Apart from this restriction, we will be able to discuss non-abelian configurations obeying the first two out of the three aforementioned algebras in full generality. We derive a general criterion that the AdS vacua must respect in order to admit $\mathfrak{su}(2)\oplus\mathbb{R}^{6-p}$ or $\mathfrak{su}(2)\oplus\mathfrak{su}(2)\oplus\mathbb{R}^{3-p}$ non-abelian branes which are energetically favored over their abelian counterpart.

In particular, specializing our results to the case of D$(d-2)$ domain-wall branes embedded in AdS$_d$ vacua, we calculate their charge and tension. Their charge is the same as the one of their abelian counterparts, confirming the dielectric nature of these bound states; their tension can be lowered or increased via the non-abelian contribution. 
In this case, our criterion to admit $\mathfrak{su}(2)\oplus\mathbb{R}^{6-p}$ or $\mathfrak{su}(2)\oplus\mathfrak{su}(2)\oplus\mathbb{R}^{3-p}$ branes takes the form of a bound involving the fluxes, and the Hessian of a combination of the warp factor and the dilaton. 
The vacua satisfying this criterion admit non-abelian branes that are less self-attractive than their abelian counterpart; they could therefore be destabilised by the former even if they resist decays mediated by the latter. This potentially opens up a new decay channel. However, since the non-abelian contributions must be small, this mechanism is only relevant when the corresponding abelian branes are close to extremality (or extremal).

This sheds interesting light on supersymmetric AdS vacua. If these have extremal abelian D$(d-2)$ domain-wall branes, they can't develop their superextremal non-abelian cousins, which places constraints on the background fluxes. To illustrate, for vacua without warping and with trivial dilaton profile, this requires the absence of $H$-flux. As the presence of an $H$-flux prevents AdS vacua to be parity symmetric, this somewhat resonates with the result that the extremality of abelian domain-walls is spoiled by corrections for non-parity-protected $\mathcal{N}=1$ AdS vacua \cite{Montero:2024qtz}. Alternatively, if such a supersymmetric solution of supergravity has both extremal abelian D$(d-2)$ domain-wall branes and a non-vanishing $H$-flux, this could indicate that the non-abelian stringy corrections actually prevent the vacua to be supersymmetric in full string theory.

We investigate two distinct types of non abelian D$(d-2)$ domain-wall branes: the ones where the non-commuting scalars are all internal, and those where one of these scalars is along the radial direction of AdS. We refer to these branes as having internal and radial fuzziness, respectively. These two types of branes have contrasting properties. In particular, branes with radial fuzziness can exist only if their abelian counterpart is superextremal; branes with purely internal fuzziness are free from this constraint. From a vacuum destabilization perspective, this makes the radially fuzzy branes less interesting than the internally fuzzy ones. However, some of these non-abelian solutions are energetically favored over the abelian branes, and are thus the ones actually triggering the decay.

We illustrate our results by embedding non-abelian domain-wall D-branes into various AdS vacua. We discuss both $\mathfrak{su}(2)\oplus\mathbb{R}^{6-p}$ and $\mathfrak{su}(2)\oplus\mathfrak{su}(2)\oplus\mathbb{R}^{3-p}$ branes, with both radial and purely internal fuzziness. We destabilise some vacua that were resisting the abelian decay channels, in particular some AdS$_4\times \mathbb{CP}^3$ and AdS$_4\times \mathbb{F}(1,2;3)$ vacua, whose abelian non-perturbative stability has been studied in 
\cite{Menet:2025nbf}.
\section{Non-abelian branes and flux vacua}

\subsection{Myers' action}\label{Myers}

We quickly present the Dirac-Born-Infeld (DBI) and Chern-Simons (CS) action proposals of Myers' \cite{Myers:1999ps} for a stack of $N$ coincident D$p$-branes.

We work in the string frame. The ten-dimensional coordinates with indices $M=0,...,9$ split into the world-volume coordinates $\sigma^a$, $a=0,\dots,p$, and the transverse coordinates $x^i$, $i= p+1,\dots,9$. In the static gauge, the non-abelian transverse scalars are $\Phi^i(\sigma)$, defined through the transverse coordinates as $x^i=\lambda \Phi^i$, with the usual $\lambda= 2\pi l_s^2.$ They are matrix-valued in the adjoint of the U$(N)$ group defined by the stack of N coincident D$p$-branes, and they are of dimension $length^{-1}$. The world-volume gauge field $A_a$ is similarly matrix-valued, with field-strength $\mathcal{F}_{ab}=\partial_a A_b-\partial_b A_a + i[A_a,A_b]$.

Myers' proposal for the non-abelian DBI action is
\begin{equation}\label{DBI}
S_{\text{DBI}} = -T_p\int d^{p+1}\sigma\;\mathrm{STr}\Bigg( e^{-\phi}\sqrt{ -\det\big( P\left[ E_{ab} + E_{ai}(Q^{-1}-\delta)^{ij}E_{jb}\right] + \lambda \mathcal{F}_{ab}\big) \det (Q^i{}_j)}\;\Bigg),
\end{equation}
where
\begin{align}
E_{MN} &= g_{MN}+B_{MN}, \\
Q^i{}_j &= \delta^i{}_j + i\lambda\,[\Phi^i,\Phi^k]E_{kj},
\end{align}
and $P[\cdot]$ denotes the pull-back to the world-volume. For instance, for a spacetime tensor $T_{MN}$, it yields
\begin{equation}
P[T]_{ab}=T_{ab}+2\lambda T_{i(a}D_{b)} \Phi^i + \lambda^2 T_{ij} D_a \Phi^i D_b \Phi^j + \cdots,\label{pullback}
\end{equation}
where the covariant derivative of transverse scalars is
\begin{equation}
D_a \Phi^i = \partial_a \Phi^i + i[A_a, \Phi^i].
\end{equation} 
The symmetrized trace STr is over gauge indices and symmetrizes over the field-strength $F_{ab}$, the commutators $[\Phi^i,\Phi^j]$, the D$_a\Phi^i$ terms and the scalars themselves. 

Generically, the fields entering the DBI action depend on all coordinates, and as such are functions of the non-abelian scalars. They must therefore be Taylor expanded about the D$p$ branes stack center of mass, defined as $\{x^i=0\}$. Denoting with $|$ the pull-back to this locus, again for a spacetime tensor $T_{MN}$:
\be T_{MN}=\exp[\lambda\Phi^i\partial_{x_i}]T_{MN}|.\label{Taylor}\ee
Notice the difference of $|$ with $P$ in \eqref{pullback}.
For a poly-form $\alpha$, we define $\alpha|$ to be the pull-back of the $p+1$-form part $\alpha_{p+1}$ to $\{x^i=0\}$.

    Crucially, the DBI action is only trustworthy up to order $\alpha'^2$, as it reproduces this order open-string amplitudes results \cite{Bergshoeff:2001dc}, but fails at higher order \cite{Bilal:2001hb}. This is because the symmetrized trace prescription isn't the correct structure to account for higher order contributions. There is no known closed form expression of the DBI action to all orders in $\alpha'$.

The non-abelian Chern-Simons action reads
\begin{equation}\label{CS}
S_{\text{CS}} = \mu_p \int \mathrm{STr}\left(P\left[ e^{i\lambda\,\iota_\Phi\iota_\Phi}\!\left( C\wedge e^{B}\right) \right] e^{\lambda \mathcal{F}}\right),
\end{equation}
 where $\iota_\Phi$ is the interior product with $\Phi$ here thought off as a transverse vector. For instance, for a two-form potential $C^{(2)}$, we have
\begin{equation}
\iota_\Phi\iota_\Phi\,C^{(2)}=\Phi^i\Phi^jC^{(2)}_{ji}=\frac{1}{2}[\Phi^i,\Phi^j] C^{(2)}_{ji}.
\end{equation}
The RR and NS fields must again be Taylor expanded about $x^i=0$ as in \eqref{Taylor}.
As for the DBI action, this action can be trusted only up to order $\alpha'^2$. 

The Myers action must be understood as the action of a stack expanded around a stationary point of the abelian potential 
\be
\partial_iV_\mathrm{abelian}=0,\label{min}
\ee 
with $V_\mathrm{abelian}=T_pe^{-\phi}\sqrt{-\mathrm{det}P[E]_{ab}}\dd^{p+1}\sigma-\mu_pP[C\wedge e^B]$. Additionally, the branes sit at an abelian minimum if $\partial_i\partial_jV_\mathrm{abelian}>0$. If they don't, the center of mass mode of the stack is tachyonic.

The charge and tension of a single D$p$-brane are
\be T_p=\mu_p=\frac{2\pi}{g_s(2\pi l_s)^{p+1}}.\ee

Given that the effective action for non-abelian branes can only be trusted up to order $\alpha'^2$, it is fundamental to work in the regime where said action can be sensibly truncated to this order.
There are two expansions at play in the Myers action: the non-abelian one with an expansion parameter going like $\sim\lambda[\Phi,\Phi]$, and the Taylor expansion about the stack center of mass, with $\lambda\Phi^i\partial_{x^i}$ as an expansion parameter. Both expansions contribute as small corrections to the abelian case if 
    \be \lambda\sqrt{\frac{1}{N}\mathrm{Tr}([\Phi^i,\Phi^j][\Phi^j,\Phi^i])}\ll 1,\qquad\frac{\lambda}{L}\sqrt{\frac{1}{N}\mathrm{Tr}(\Phi^i\Phi^i)}\ll 1, \label{expansion2}\ee
    where $L$ is a typical length. We stay within this regime throughout this paper, and truncate the actions \eqref{DBI} and \eqref{CS} to the terms of order up to $\alpha'^2$, so corresponding to a second order term in either expansion, or a combination of their first order expansions.
    
    It is also important to note that the Myers action is derived for slowly varying background fields, in the sense that the matching with the string amplitude action has been carried out discarding the terms with more than two derivative. However, in generic curved flux backgrounds, there can be $\alpha'^2$ four-derivative contributions: one must ensure that they are negligible when compared to the non-abelian contributions so that the Myers action is indeed a sensible effective action at this order.
Schematically, the abelian four-derivative $\alpha'^2$ terms go like $\sim\alpha'^2 R^2$ and $ \alpha'^2(\partial H)^2$\cite{Bachas:1999um,Garousi:2009dj}, with $R$ a placeholder for Riemann and Ricci tensors.\footnote{There are additional dilaton derivative terms, terms mixing these contributions, but they all have the same parametric size. There are also derivative terms in the open string sector like $\alpha'^2(D\mathcal{F})^2$, but one usually takes the variation of world-volume fields to be parametrically slow.} Their non-abelian counter parts are not known, but matching with the abelian limit they should at least include terms like $\sim\alpha'^2 N R^2$ and $ \alpha'^2 N (\partial H)^2$.

We will come back to this issue more concretely when focusing on some specific non-abelian configurations, and compare the parametric size of these contributions and those of Myers'.

\subsection{General expansion of the Myers action}

The Myers action has been used to unravel the celebrated Myers effect of dielectric branes \cite{Myers:1999ps}, which was illustrated in flat space and with a single RR three-form potential. This simple configuration brings the effective action of a stack of $N$ coincident D$0$-branes to a tractable form with a straightforward equation of motion for the non-abelian scalars, solved for representations of an $\mathfrak{su}(2)$ algebra. 

We will now analyze the Myers action more generally, expanding around a general brane in curved space. Our only simplifying assumption will be that on the branes:
\begin{equation}
    E_{ai}|=0 \,.   
\end{equation}
Luckily, there is no real loss of generality in assuming this. We can set $g_{ai}|=0$ on the brane by choosing Fermi coordinates adapted to it (see for example \cite[Prop.~5.26]{lee-riemannian}). Moreover, if $B_{ai}|\neq 0$, the gauge transformation $B\to B+ \dd \lambda$, $\lambda= x^i B_{ai}\dd \sigma^a$ can be used to set $B_{ai}|=0$ on the brane, where $x^i=0$. 

We introduce the notation
\begin{equation}
	K^a{}_{bi} \equiv \tE^{ac} \partial_i \tE_{cb} \, ,\qquad K_i \equiv K^a{}_{ai}\,.
\end{equation}
When $B=\phi=0$, $K^a{}_{bi}$ is proportional to the second fundamental form.

We set $\Tilde E_{ab}=e^{-2\phi/(p+1)}E_{ab}$. Expanding $Q$ and Taylor expanding $\Tilde E$, 
a lengthy but straightforward computation gives for the DBI action \eqref{DBI}, up to order $\alpha'^2$:
\begin{align} \label{eq:DBI-exp}
	S_\mathrm{DBI}=&-T_{p}\int \dd^{p+1}\sigma\;\sqrt{ -\det(\Tilde E_{ab})}\,\mathrm{STr}\left[\mathbb{I}
	+\frac{\lambda}{2}(E^{ac}\mathcal{F}_{ca}+ \Phi^i K_i)\right.
	\nonumber\\
&+\lambda^2\left[-\frac14 E^{ac} \mathcal{F}_{cb} E^{bd}\mathcal{F}_{da} +\frac18 (E^{ac}\mathcal{F}_{ca})^2 + \Phi^i\left(-\frac12 E^{ac} \mathcal{F}_{cb} K^b{}_{ai}+\frac14 K_i E^{bd}\mathcal{F}_{db}\right)\right.\nonumber\\
&+\Phi^i\Phi^j\left(\frac{1}{2}\tE^{ac}\partial_i\partial_j\tE_{ca}-\frac14K^a{}_{bi}K^b{}_{aj}+\frac18 K_i K_j\right)\\
&+ \frac12 \Phi^i D_a \Phi^j ( \tE^{ac} \partial_i \tE_{cj} + \tE^{ca} \partial_i \tE_{jc} )+ \frac12 D_a \Phi^i D_b \Phi^j \tE^{ba} \tE_{ij} \nonumber\\
&\left.\left.+i \Phi^i \Phi^j\Phi^k \left( -\frac12 B_{[ij} K_{k]}-\frac13 H_{ijk}\right)+\frac{1}{4}[\Phi^i,\Phi^j][\Phi^k,\Phi^l]\tilde g_{jk} \tilde g_{li}\right]\right]\,. \nonumber
\end{align}
In the quartic term, we have used $ E_{j[k}E_{l]i}- E_{i[k}E_{l]j} -B_{ij}B_{kl}= g_{j[k}g_{l]i}- g_{i[k}g_{l]j} -3 B_{[ij}B_{kl]}$.

We now turn to the CS action. At first we compute
\begin{align}\label{eq:CS-0}
    \begin{split}
    S_\mathrm{CS}=T_{p}\int \mathrm{STr}&\bigg[ \hat C+\lambda (\Phi^i \partial_i \hat C + D \Phi^i \wedge \iota_i \hat C)
    + \lambda^2 \left(\frac12 \Phi^i \Phi^j \partial_i \partial_j \hat C +i \Phi^i \Phi^j \Phi^k \iota_i \iota_j \partial_k \hat C\right.\\
    &\left.+ D \Phi^k \wedge (\Phi^i \iota_k \partial_i \hat C+ i \Phi^i \Phi^j \iota_i \iota_j \iota_k \hat C) - \frac12 D \Phi^i \wedge D \Phi^j \iota_i \iota_j \hat C\right)\bigg],
    \end{split}
\end{align}
with $\hat C \equiv C \wedge e^B$, again up to order $\alpha'^2$. From now on, the integrands of actions are understood as being evaluated at the branes center of mass $\{x^i=0\}$; moreover, only the top-form $( \, )_{p+1}$ part of forms is kept.  

We now use $\{ \iota_i ,\dd \} = \partial_i$ repeatedly to obtain the identities
\begin{subequations}
\begin{align}
	\mathrm{Tr}\dd(\Phi^i \Phi^j \Phi^k \iota_i \iota_j \iota_k \hat C)&= \mathrm{Tr}\left(\Phi^i \Phi^j(3 \dd\Phi^k \wedge \iota_i \iota_j \iota_k \hat C+ \Phi^k (-\iota_i \iota_j \iota_k \dd +3 \iota_{[i}\iota_j \partial_{k]} )\hat C)\right)\,,\\
	\mathrm{Tr}D(\Phi^i D \Phi^j\wedge \iota_i \iota_j \hat C)&= 
	\mathrm{Tr}\left((D \Phi^i \wedge D \Phi^j +\Phi^i [\mathcal{F},\Phi^j] )\wedge \iota_i \iota_j \hat C - \Phi^i D \Phi^j \wedge \dd \iota_i \iota_j \hat C\right)\,,\\
	\mathrm{Tr}D(\Phi^i \Phi^j \iota_i \partial_j \hat C) &= \mathrm{Tr}(2 D \Phi^i \Phi^j \iota_{(i} \partial_{j)} \hat C+ \Phi^i \Phi^j (-\iota_i\dd + \partial_i ) \partial_j\hat C)
\end{align}
\end{subequations}

Combining these and discarding total derivatives, we can rewrite \eqref{eq:CS-0} as
\begin{align}\label{eq:CS-1}
    \begin{split}
    S_\mathrm{CS}=T_{p}\int \mathrm{STr}\bigg[ \hat C+\lambda \Phi^i \iota_i \hat F
    &+ \lambda^2 \left(\frac12 \Phi^i \Phi^j  (\iota_i \partial_j \hat F - \mathcal{F} \wedge \iota_i\iota_j \hat C ) + \right.\\
    &\left. +\frac12 \Phi^i D \Phi^j \wedge \iota_i \iota_j \hat F + \frac i3 \Phi^i \Phi^j \Phi^k \iota_i \iota_j \iota_k \hat F\right)\bigg]\,,
    \end{split}
\end{align}
where we have introduced $\hat{F}=\dd\hat{C}$ and discarded a total derivative.

Crucially, both actions are only sensible when the branes sit at an extremum of the abelian action. We revisit this more concretly in the next section.

\subsection{Constant transverse scalars}

We will now make the further simplifying assumption
\be \partial_a\Phi^i=0 \qquad A_a=0.
\label{gencond}
\ee

\eqref{eq:DBI-exp} now becomes
\begin{align} S_\mathrm{DBI}=&-T_{p}\int \dd^{p+1}\sigma\;\sqrt{ -\det(\Tilde E_{ab})}\,\mathrm{STr}\left(\mathbb{I}+\frac{\lambda}{2}\Phi^i\tE^{ac}\partial_i \tE_{ca}\right. \nonumber\\
&\left.+\lambda^2\left[\Phi^i\Phi^j\left(\frac{1}{2}\tE^{ac}\partial_i\partial_j\tE_{ca}-\frac14\tE^{ac}\partial_i\tE_{cb}\tE^{bd}\partial_j\tE_{da}+\frac18\tE^{ac}\partial_i\tE_{ca}\tE^{bd}\partial_j\tE_{bd}\right)\right.\right.\nonumber\\
&\left.\left.-[\Phi^i,\Phi^j]\Phi^k\left(\frac{i}{4}B_{[ij}\tE^{ac}\partial_{k]}\tE_{ca}-\frac i6H_{ijk}\right)+\frac{1}{4}[\Phi^i,\Phi^j][\Phi^k,\Phi^l]\Tilde{g}_{jk}\Tilde{g}_{li}\right]\right)\,.
\end{align}
 The CS action \eqref{eq:CS-1} simplifies to
\begin{align}
    S_\mathrm{CS}=&T_{p}\int \mathrm{STr}\bigg[ \hat{C}+\lambda \Phi^i\iota_i\hat{F}+\frac{\lambda^2}{2}\bigg(\Phi^i\Phi^j\iota_i\partial_j\hat{F}+ \frac i6[\Phi^i,\Phi^j]\Phi^k\iota_{[i}\iota_j\iota_{k]}\hat{F}\bigg)\bigg].
\end{align} 

As we mentioned above, these actions are only sensible when the branes sit at an extremum of the abelian action, that is \eqref{min} must be respected. We set $\dd\widetilde{\mathrm{vol}}=\dd^{d-1}\sigma\;\sqrt{ -\det(\Tilde E_{ab})}$, and this entails 
\be  \frac{1}{2}\dd\widetilde{\mathrm{vol}}\;\tE^{ac}\partial_i \tE_{ca}=\iota_i\hat{F}|.\label{min2}
\ee
The combined action then reduces to
\begin{align}
    S&=S_\mathrm{DBI}+S_\mathrm{CS}\nonumber\\
    &=T_{p}\int  N( \hat{C} -\dvol)+\dvol\lambda^2\mathrm{STr}\bigg(-\frac12\Phi^i\Phi^jS_{ij}\nonumber\\
    &\quad+\Phi^i[\Phi^j,\Phi^k]\left(\frac{i}{6\dvol}(3B_{[ij}\iota_{k]}+\iota_{[i}\iota_j\iota_{k]})\hat{F}+\frac i6H_{ijk}\right)-\frac{1}{4}[\Phi^i,\Phi^j][\Phi^k,\Phi^l]\tilde g_{jk} \tilde g_{li}\bigg),\label{Sgen}
\end{align}
with
\begin{align}
S_{ij}&=\tE^{ac}\partial_i\partial_j\tE_{ca}-\frac12\tE^{ac}\partial_i\tE_{cb}\tE^{bd}\partial_j\tE_{da}+\frac14\tE^{ac}\partial_i\tE_{ca}\tE^{bd}\partial_j\tE_{bd}\nonumber\\
&\quad-\frac{1}{\dvol}\iota_i\partial_j\hat{F}|.\label{sij}
\end{align}
The stack sits at an abelian minimum if $S_{ij}>0$. 

The presence of the quadratic $S_{ij}$ represents one of the novelties of our approach. The metric terms can be interpreted in terms of ambient curvature, but we will not need this in what follows.
We will see shortly that we are able to solve the equations of motion under some conditions on the $S_{ij}$.

The equations of motion for the scalars $\delta S/\delta \Phi^i=0$ are 
\be0=[\Phi^j,[\Phi^k,\Phi^l]]g_{jk}g_{li}
-\frac i2[\Phi^j,\Phi^k]\left(\frac{1}{\dvol}(3B_{[ij}\iota_{k]}+\iota_{[i}\iota_j\iota_{k]})\hat{F}|+H_{ijk}\right)
+\Phi^jS_{ij}.\label{eomgen}\ee
We now investigate the possible Lie algebras satisfied by the scalars that might be solution to these equations. 

\subsection{A simple solution}
\label{sub:simple}

The scalars are finite-dimensional representations of a $\mathfrak{u}(N)$ subalgebra, and as such their hermiticity requires their algebra to be reductive. The candidate algebras are therefore
\be \mathfrak{su}(2)\oplus\mathbb{R}^{6-p},\qquad \mathfrak{su}(2)\oplus\mathfrak{su}(2)\oplus\mathbb{R}^{3-p}\qquad\mathfrak{su}(3)\oplus\mathbb{R}^{1-p},\ee
if $p\leq6,3,1$ respectively.

\subsubsection{\texorpdfstring{The $\mathfrak{su}(2)$ branes}{The su(2) branes}}

Let us focus on the $\mathfrak{su}(2)\oplus\mathbb{R}^{6-p}$ possibility, and keep only three non-vanishing scalars. On the three-dimensional semi-simple subspace, we set 
\begin{subequations}\label{3d}
\begin{align}
    f\epsilon_{ijk}&=\frac{1}{\dvol}(3B_{[ij}\iota_{k]}+\iota_{[i}\iota_j\iota_{k]})\hat{F}|\\
    h\epsilon_{ijk}&=H_{ijk}.
\end{align}
\end{subequations}
This choice of parametrisation is simply the most convenient to describe domain-wall branes, as we will see in the next section. The equations of motion are 
\begin{align} 
[\Phi^j,[\Phi^k,\Phi^l]]g_{jk}g_{li}
-i\frac {f+h}{2}[\Phi^j,\Phi^k]\epsilon_{ijk}+\Phi^jS_{ij}=0.\label{eomphigen}
\end{align}

We consider the possibility that the scalars obey an algebra isomorphic to $\mathfrak{su}(2)$:
\be [\Phi^i,\Phi^j]=-i\frac{f+h}{2}{M^i}_p{M^j}_q{\epsilon^{pq}}_r{(M^{-1})^r}_k\Phi^k\label{isosu2}\ee
for $M$ some invertible matrix. We set $P_{ij}=\mathrm{det}M((M^{-1})^TgM^{-1})_{ij}$ and $t=\mathrm{Tr}P$. This implies that the eigenvalues $p_i$ of $P$ are either all positive or all negative. One can relate the two cases by a permutation or by changing sign to all the $\Phi^i$; thus in the following we restrict to the case where all the $p_i > 0$.

Before studying the solutions to the equations of motion, let us briefly discuss the action \eqref{Sgen} for branes satisfying the algebra \eqref{isosu2}: it becomes
 \begin{align} S&=T_{p}\int  N(\hat{C}-\dvol)+\dvol\lambda^2\frac{(f+h)^4}{3\cdot 2^7}\frac{\mathrm{det}P}{\mathrm{det}g_{\mathfrak{su}(2)}}N(N^2-1)(t-2),\label{actionsu2gen}
\end{align}
where we used the usual spin-$j$ $\mathfrak{su}(2)$ irrep with $N=2j+1$, and with $g_{\mathfrak{su}(2)}$ the pull-back of the metric on the three-dimensional semi-simple subspace. The non-abelian potential is therefore
\be V=\lambda^2\frac{(f+h)^4}{3\cdot 2^7}\frac{\mathrm{det}P}{\mathrm{det}g_{\mathfrak{su}(2)}}N(N^2-1)(2-t).\ee
The potential is therefore negative when $t>2$: in that case, the non-abelian stack is then energetically favored compared to its abelian counterpart. We therefore focus on $t>2$, and the equations of motion \eqref{eomphigen} become
\be \label{eq:SP}
\frac{4}{(f+h)^2}S_{ij}-2P_{ij}-P_{ki}{P^k}_j+tP_{ij}=0.
\ee
When these equations are satisfied, the triplet of scalars depends on the three-dimensional matrix $S_{ij}$ via $M_{ij}$. Since we are working under the assumption that they are constant along the world-volume directions, the requirement \eqref{gencond} imposes 
\be\partial_aS_{ij}=0.\label{condS}\ee
We will come back to this condition more concretely in the next section.

Since $P$ is symmetric, it can be considered diagonal, $P=\textrm{diag}(p_1,p_2,p_3)$, in the basis that diagonalises $S$. \eqref{eq:SP} then give $p_i = \frac12(t-2\pm\sqrt{(t-2)^2+16 (f+h)^{-2}s_i})$, and for $t>2$ there is always at least one positive branch.

There are three distinct cases: when all the $s_i$ are non-negative, when one of the $s_i$ is negative, and finally when two or all of the $s_i$ are negative.
\begin{itemize}
\item We start with the first case, $s_i\ge 0$. We must take $\pm=+$ in the expression for the $p_i$. Recalling $t=\sum_i p_i$, they yield
\be 
f(t)\equiv t-6+\sum_{i=1}^3\sqrt{(t-2)^2+\frac{16s_i}{(f+h)^2}}=0\,.
\label{su2root}\ee
This equation only admits solutions for $t<6$, so we consider here the interval $t\in[2,6[$, for which $f(t)$ increases monotonically. Since $f(6)>0$, the equations of motion have a solution with $t>2$ if and only if $f(2)\leq0$, which yields
\be \sum_{i=1}^3\sqrt{\frac{s_i}{(f+h)^2}}\leq1.\label{su2condgen}\ee
This is a general requirement on any flux vacua to admit an $\mathfrak{su}(2)\oplus\mathbb{R}^{6-p}$ stack of branes that sits at an abelian minimum. 

\item We move on to the second case: when one of the $s_i$ is negative, say $s_3$, a similar reasoning leads to
\be  \sum_{i=1,2}\sqrt{\frac{s_i-s_3}{(f+h)^2}}\leq1-\sqrt{\frac{-s_3}{(f+h)^2}}\label{su2condgenneg}\ee
being a sufficient criterion for the equations of motion to admit at least one solution with $t>2+4\sqrt{\frac{-s_3}{(f+h)^2}}$. 

\item Finally, when two or all of the $s_i$ are negative, the equations of motion always have at least one solution.

\end{itemize}

We will come back to these conditions more concretely when considering more specific branes set up.
 
When the conditions \eqref{su2condgen}, \eqref{su2condgenneg} aren't satisfied, the flux vacua can still possibly admit an $\mathfrak{su}(2)\oplus\mathbb{R}^{7-d}$ stack with $t\in[0,2]$, these will simply be more energetic than their abelian counterpart.

Now that we consider a specific algebra satisfied by the stack, let us come back to assessing how well-controlled the truncation to $\alpha'^2$ is, and compare its parametric size with the one of the other possible $\alpha'^2$ contributions. We do so for the case of $M=\mathbb{I}$ for simplicity, with a similar behaviour for more general $M$.

Considering $h,\,f\sim 1/L$, the parametric size of the non-abelian contribution to this action goes like $\sim \left(\frac{l_s}{L}\right)^4N(N^2-1)$. In order to safely truncate the action to this order, we must satisfy \eqref{expansion2}, which here collapse into
\be \left(\frac{l_s}{L}\right)^2N\ll 1. \label{boundsu2}\ee
The $\alpha'^2$ four-derivative terms going beyond this non-abelian discussion, mentioned at the end of the section \ref{Myers}, have parametric size $\left(\frac{l_s}{L}\right)^4N$. Myers' non-abelian terms thus dominate the effective action at order $\alpha'^2$ if
\be N^2\gg1.\label{alpha2}\ee
$N$ must therefore be taken to be large for the non-abelian contribution to dominate, but not so large that it breaks down the truncation of Myers' action \eqref{boundsu2}. How large it is allowed to be is determined by two quantities. First the $\frac{l_s}{L}$ ratio via \eqref{boundsu2}, it is small in the supergravity regime. Second, the string coupling $g_s$: the back-reaction of the probe stack is in check if \be g_s N\ll \left(\frac{L}{l_s}\right)^{7-p},\ee which also caps the allowed value for $N$.

Hence, our construction reliably describes non-abelian stacks of D-branes for a range of brane numbers determined by $\frac{l_s}{L}$ and $g_s$. It may be challenging to tune $g_s$ and  $\frac{l_s}{L}$ to extremely small values, but through \eqref{alpha2} we have that $N=10$ already suffices for \eqref{actionsu2} to be a sensible effective action for our non-abelian stack.

\subsubsection{\texorpdfstring{The $\mathfrak{su}(2)\oplus\mathfrak{su}(2)$ branes}{The su(2)+su(2) branes}}\label{secsu2su2}

We focus here on $d\leq4$, and consider the non-abelian stack of branes obeying an $\mathfrak{su}(2)\oplus\mathfrak{su}(2)\oplus\mathbb{R}^{3-p}$ algebra, so keeping six non-vanishing scalars.

We divide them into two independent families living on three-dimensional subspaces $[\Phi^\alpha,\Phi^A]=0$, with the transverse indices splitting as $i=A,B,C,\alpha,\beta,\gamma$. On each three-dimensional subspace, we set
\begin{subequations}\label{su2su2param}
\begin{alignat}{2}
    f_1\epsilon_{ABC}&=\frac{1}{\dvol}(3B_{[AB}\iota_{C]}+\iota_{[A}\iota_B\iota_{C]})\hat{F}|\qquad &&h_1\epsilon_{ABC}=H_{ABC}\\
   f_2\epsilon_{\alpha\beta\gamma}&=\frac{1}{\dvol}(3B_{[\alpha\beta}\iota_{\gamma]}+\iota_{[\alpha}\iota_\beta\iota_{\gamma]})\hat{F}|\qquad  &&h_2\epsilon_{\alpha\beta\gamma}=H_{\alpha\beta\gamma}.
\end{alignat}
\end{subequations}
The equations of motion for the scalars \eqref{eomgen} become
\begin{align} 
0&=[\Phi^C,[\Phi^B,\Phi^D]]g_{BC}g_{DA}-i\frac{f_1+h_1}{2}[\Phi^B,\Phi^C]\epsilon_{ABC}+\Phi^BS_{AB}\nonumber\\
&\quad-\frac{i}{2}[\Phi^\alpha,\Phi^\beta]\left(\frac{1}{\dvol}(3B_{[A\alpha}\iota_{\beta]}+\iota_{[A}\iota_\alpha\iota_{\beta]})\hat{F}|+H_{A\alpha\beta}\right)+\Phi^\alpha S_{A\alpha}\label{eomphi2}
\end{align}
for $\delta S/\delta \Phi^A=0$, with similar equations for $\delta S/\delta \Phi^\alpha=0$.
The first line vanishes when each set of scalars obey the appropriate $\mathfrak{su}(2)$ algebra discussed in the previous section
\begin{subequations}\label{isosu2su2gen}
\begin{align} [\Phi^A,\Phi^B]&=-i\frac{f_1+h_1}{2}{M_1^A}_P{M_1^B}_Q{\epsilon^{PQ}}_R{(M_1^{-1})^R}_C\Phi^C\\
[\Phi^\alpha,\Phi^\beta]&=-i\frac{f_2+h_2}{2}{M_2^\alpha}_\delta{M_2^\beta}_\mu{\epsilon^{\delta\mu}}_\nu{(M_2^{-1})^\nu}_\gamma\Phi^\gamma,
\end{align}
\end{subequations}
with $M_1$ and $M_2$ invertible matrices. Plugging these in, the second line of \eqref{eomphi2} vanishes if
\begin{subequations}\label{su2su2eom}
\begin{align}
4S_{D\mu}&=(f_2+h_2)\epsilon^{\alpha\beta\gamma}P_{2\gamma\mu}\left(\frac{1}{\dvol}(3B_{[D\alpha}\iota_{\beta]}+\iota_{[D}\iota_\alpha\iota_{\beta]})\hat{F}|+H_{D\alpha\beta}\right)\\
4S_{\mu D}&=(f_1+h_1)\epsilon^{ABC}P_{1CD}\left(\frac{1}{\dvol}(3B_{[\mu A}\iota_{B]}+\iota_{[\mu}\iota_A\iota_{B]})\hat{F}|+H_{\mu AB}\right)
\end{align}
\end{subequations}
with $P_{1AB}=\mathrm{det}M_1((M_1^{-1})^TgM_1^{-1})_{AB}$ and $P_{2\alpha\beta}=\mathrm{det}M_2((M_2^{-1})^TgM_2^{-1})_{\alpha\beta}$. 
We don't study general solutions to these equations of motion, and consider instead the case where
\begin{subequations}\label{triveom}
\begin{align}&S_{A\alpha}=(3B_{[A\alpha}\iota_{\beta]}+\iota_{[A}\iota_\alpha\iota_{\beta]})\hat{F}|=H_{A\alpha\beta}=0\\
&S_{\alpha A}=(3B_{[\alpha A}\iota_{B]}+\iota_{[\alpha}\iota_A\iota_{B]})\hat{F}|=H_{\alpha AB}=0,
\end{align}
\end{subequations}
such that the equations of motion \eqref{su2su2eom} are trivially satisfied. In the next section we present vacua with domain-wall branes satisfying these conditions.

    Again using the usual spin-$j$ $\mathfrak{su}(2)$ irrep, the non-abelian potential for these $\mathfrak{su}(2)\oplus\mathfrak{su}(2)\oplus\mathbb{R}^{3-p}$ branes in vacua satisfying \eqref{triveom} is
 \begin{align} V=\frac{\lambda^2}{3\cdot 2^7}N(N^2-1)\left(\frac{\mathrm{det}P_1}{\mathrm{det}g_1}(f_1+h_1)^4(2-t_1)+\frac{\mathrm{det}P_2}{\mathrm{det}g_2}(f_2+h_2)^4(2-t_2)\right),\label{potsu2su2gen}
\end{align}
with $\mathrm{det}g_{1,2}$ the pull-back on the corresponding three-dimensional subspace, and with $t_{1,2}=\mathrm{Tr}P_{1,2}$. This potential simply contains the two $\mathfrak{su}(2)$ contributions.

    The regime of validity is as for the $\mathfrak{su}(2)$ case
\be N^2\gg1\qquad g_s N\ll \left(\frac{L}{l_s}\right)^{7-p}\qquad \left(\frac{l_s}{L}\right)^2N\ll1.\label{regimesu2su2}\ee

We witness the same behaviour as for the $\mathfrak{su}(2)$ case: when the $s_{1i}$, $s_{2i}$ are positive and the conditions
\be \sum_{i=1}^3\sqrt{\frac{s_{1i}}{(f_1+h_1)^2}}\leq1,\qquad\sum_{i=1}^3\sqrt{\frac{s_{2i}}{(f_2+h_2)^2}}\leq1\label{su2su2condgen}\ee
are satisfied, the stack is energetically favored compared to its abelian counterpart. Here $s_{1i}$, $s_{2i}$ are the eigenvalues of $S_{AB}$, $S_{\alpha\beta}$ respectively. If both conditions fail, the possible solutions to the equations of motion have both $t_1<2$ and $t_2<2$, and the abelian branes are favored. If only one of these conditions is met, the situation gets decided case by case by the competition of both contributions.
\subparagraph{$\mathfrak{su}(3)$ branes} 
We briefly discuss $\mathfrak{su}(3)$ branes in the next section within the context of domain-wall branes.

\section{Purely domain-wall branes}
We now apply the results of the previous section to type II AdS$_d\times M_{10-d}$ warped vacua with metric
\be 
\label{10dmet}
\dd s^2_{10}=e^{2A}\dd s^2_{\rm AdS_d} + \dd s^2_{10-d}\, ,
\ee
and we will consider throughout this section the case of domain-wall D$(d-2)$ branes. We first quickly analyse the stability of the vacua probed by such branes, from the effective perspective, before moving on to discussing the specifics of non-abelian domain-walls. 
\subsection{Domain-wall instability}
\label{sub:stab}
In the Wick-rotated Euclidean AdS$_d$ vacua with metric
\begin{equation}\label{eq:eads}
    \dd s^2_{\rm EAdS}= L^2 (\dd r^2 +\sinh^2 r \dd s^2_{S^{d-1}})\,,
\end{equation}
a vacuum decay can be triggered by a localised instantonic bubble of new vacuum at fixed $r$ and along $S^{d-1}$\cite{Coleman:1980aw}.
Let us consider this bubble to be a probe brane with action
\be S=-\tau \int \dd^{d-1} \sigma \sqrt{-g}- q \int A_{d-1},\ee
with $A_{d-1}$ a $(d-1)$-dimensional gauge field.
Its field-strength is proportional to the volume form: $F_d = f {\rm vol}_d$. Specializing the Wick-rotated action to the metric \eqref{eq:eads} yields \cite{Apruzzi:2019ecr}
\begin{equation}\label{eq:s-inst}
    S= -L^{d-1} {\rm Vol}(S^{d-1}) (\tau \sinh^{d-1} r  + q f 
    \, c(r)),
\end{equation}
where $c'(r)= \sinh^{d-1}r$. The spherical symmetry of the brane reduces the search for the instanton to extremizing \eqref{eq:s-inst} in $r$. This gives 
\be\tanh r_0 =- \frac{(d-1) \tau}{q f}\label{instrad}\ee
for $r_0$  the radius of the instanton. The instanton, and thus the instability of the AdS vacua in which it is embedded, hence exists only if
\begin{equation}\label{eq:dq-stab}
\frac{(d-1) \tau}{|q| f}<1 \,.
\end{equation}
In the $d=3$ case, there can be a non-vanishing NS field-strength along AdS$_3$: $H=h\mathrm{vol}_3$. The corresponding $B$-field can be taken along the two-sphere with $B=h\, c(r) {\rm vol}(S^{2})$, and the brane action becomes
\begin{equation}\label{eq:s-instB}
    S= -L^{2} {\rm Vol}(S^{2}) \left(\tau\sqrt{\sinh^{4} r +h^2\; c^2(r)} + q f  \, c(r)\right) 
\end{equation}
for the Euclidean metric \eqref{eq:eads}. The instanton radius is now defined by
\be \frac{\sqrt{\sinh^4{r_0}+h^2(\cosh r_0\sinh r_0-r_0)^2/4}}{\cosh r_0 \sinh r_0 +h^2(\cosh r_0\sinh r_0-r_0)/4}=\frac{2\tau}{|q|f},\,\label{instrB}\ee
and the instability exists if
\be \frac{2\tau}{|q|f}<\frac{1}{\sqrt{1+h^2/4}}.\label{instabB}\ee
For $h=0$, this coincides with the previous result. 
\subsection{Internal fuzziness}
Coming back to non-abelian domain-wall D$(d-2)$ branes, let us first revisit our general constraints \eqref{gencond}. Since the branes we consider here are purely external, their world-volume field-strength must vanish to preserve the AdS isometries,\footnote{With the exception of the AdS$_3$ case where one could have a world-volume field-strength along $S^2$.} so in this set up the trivial gauge $A_a=0$ is a natural choice. Moreover, the compactification ansatz \eqref{10dmet} automatically ensures $E_{ai}=0$, so we trivially are in the right Fermi coordinates.

In this section we consider non-abelian brane configurations with a purely internal fuzziness, in the sense that we impose for the transverse scalar associated to the radial direction of AdS to vanish:
\be \Phi^r=0.\ee

Our conventions for the RR fluxes are
\be 
F_{10}=F+e^{dA}\text{vol}_d\wedge\ast\sigma F\label{RR},
\ee
with  $\sigma$ the reversal of all form indices: $\sigma \alpha_{m_1\ldots m_k} = \alpha_{m_k\ldots m_1} $.
In the CS action we will make use of the following gauge for the RR potentials
\be C=c_{\text{int}}+L^dc(r)\mathrm{vol}_{S^{d-1}}\wedge (e^{dA}\ast\sigma F+h\, c_{\text{int}}\,\delta_{3,d})\label{CRR},\ee
with $c_{\text{int}}$ the purely internal potential d$c_{\text{int}}=F$, and $h$ the magnitude of the NS field-strength along AdS$_3$, in the $d=3$ case. We present the cases where there is no such $H$-flux throughout this section for simplicity, as all the results can be naturally extended to include its contribution.

We define $\varphi=(d-1)A-\phi$, and the RR equations of motion \be\dd(e^{dA}\ast\sigma F\wedge e^B)=0\label{RReq}\ee simplify both the minimisation of the abelian potential \eqref{min2} to
\be \partial_i\varphi=0,\ee
and the action \eqref{Sgen} to
\begin{align} S=& L^{d-1}T_{d-2} {\rm Vol}(S^{d-1})\bigg(e^\varphi\sinh^{d-1} r\Big[-N-\frac{\lambda^2}{2}\mathrm{STr}(\Phi^j\Phi^i)S_{ij}\nonumber\\&+i\frac {\lambda^2}{6}\mathrm{STr}([\Phi^i,\Phi^j]\Phi^k)H_{ijk}-\frac{\lambda^2}{4}\mathrm{STr}([\Phi^i,\Phi^j][\Phi^k,\Phi^l])g_{jk}g_{li}\Big]+Lc(r)e^{dA}N(\ast\sigma F)_0\bigg),\label{Sint}
\end{align}
with 
\be S_{ij}=\frac{2}{d-1}\partial_i\partial_j\varphi.\ee
We see here that the condition \eqref{condS}, imposed by the requirement that the scalars must be independent of the world-volume directions, is indeed satisfied.
The equations of motion \eqref{eomgen} reduce to
\be0=[\Phi^j,[\Phi^k,\Phi^l]]g_{jk}g_{li}
-\frac i2[\Phi^j,\Phi^k]H_{ijk}
+\Phi^jS_{ij}.\label{eomint}\ee
Let us now revisit the possible Lie algebras satisfied by the scalars that might be solution to these equations. As we discussed in the previous section, the only candidate algebras are reductive, so for AdS$_d$ vacua they are:
\begin{center}
\begin{tabular}{ c | c }
  & Candidate algebras  \\ 
  \hline
 $d=2$ & $\mathfrak{su}(2)\oplus\mathbb{R}^{5},\quad 
 \mathfrak{su}(2)\oplus\mathfrak{su}(2)\oplus\mathbb{R}^{2},\quad \mathfrak{su}(3)$ \\  
 $d=3,4$ & $\mathfrak{su}(2)\oplus\mathbb{R}^{7-d},\quad \mathfrak{su}(2)\oplus\mathfrak{su}(2)\oplus\mathbb{R}^{4-d}$\\
 $d=5,6,7$ & $\mathfrak{su}(2)\oplus\mathbb{R}^{7-d}$.
\end{tabular}
\end{center}
\subsubsection{\texorpdfstring{The $\mathfrak{su}(2)$ branes}{The su(2) branes}}\label{su2intern}
Let us focus on the $\mathfrak{su}(2)\oplus\mathbb{R}^{7-d}$ possibility, and keep only three non-vanishing scalars. The domain-wall branes with purely internal fuzziness have 
\be f=0,\qquad h\neq 0.\ee
When the $s_i$ are positive, the equations of motion therefore have a solution with $t>2$ if and only if
\be \sum_{i=1}^3\sqrt{\frac{s_i}{h^2}}\leq1,\label{su2cond}\ee
where the $s_i$ are now the eigenvalues of the three-dimensional Hessian of $\varphi$.

 Let us illustrate this result with AdS vacua having constant dilaton and warp factor, like the Freund--Rubin type vacua, vacua with cosets and homogeneous internal spaces, and so on. They have $s_i=0$, so they respect \eqref{su2cond}. They therefore solve the equations of motion with $t=3$, and as such they all admit these non-abelian branes.

We now revisit what $t>2$ entails physically. To do so, we specify the effective action for the $\mathfrak{su}(2)\oplus\mathbb{R}^{7-d}$ stack respecting \eqref{isosu2}
 \begin{align} S=& L^{d-1}T_{d-2} {\rm Vol}(S^{d-1})\Big[-Ne^{(d-1)A-\phi}\sinh^{d-1} r+Lc(r) e^{dA} N(\ast\sigma F)_0\nonumber\\
 &+\frac{\lambda^2h^4}{3\cdot 2^7}\frac{\mathrm{det}P}{\mathrm{det}g_{\mathfrak{su}(2)}}N(N^2-1)e^{(d-1)A-\phi}\sinh^{d-1} r(t-2)\Big].\label{actionsu2}
\end{align}
The charge and tension of the non-abelian stack can be read off its effective action from \eqref{eq:s-inst}.
\begin{subequations}\label{qtausu2}
\begin{align}
    qf&=T_{d-2}e^{dA} L N(\ast\sigma F)_0\\
    \tau&=T_{d-2}e^{(d-1)A-\phi} N\Big[1-\frac{\lambda^2h^4}{3\cdot 2^7}\frac{\mathrm{det}P}{\mathrm{det}g_{\mathfrak{su}(2)}}(N^2-1)(t-2)\Big].\label{tausu2}
\end{align}
\end{subequations}
The charge of the non-abelian stack is the same as in the abelian case: the CS action doesn't pick up any non-abelian contribution at this order. This is the familiar statement that the non-abelian vacua are dieletric. The tension of the stack however does acquire a non-abelian contribution. This contribution lowers the tension with respect to the abelian case if $t>2$ and increases it if $t<2$. Any flux vacua satisfying \eqref{su2cond} therefore admits an $\mathfrak{su}(2)\oplus\mathbb{R}^{7-d}$ stack of purely domain-wall branes that are less self-attractive than their abelian counterpart.

It is a very interesting feature of this non-abelian configuration: consider an abelian stack of domain-wall branes that doesn't trigger a decay of the flux vacua it is probing. It could be collapsing onto itself, or subject to a no-force condition (extremal). Its non-abelian counterpart has the same charge but a lower tension, which can bring the non-abelian stack to satisfy the instability condition \eqref{eq:dq-stab} and hence expand, opening up a new decay channel.

As discussed above, when this condition \eqref{su2cond} isn't satisfied, the flux vacua can still admit $\mathfrak{su}(2)\oplus\mathbb{R}^{7-d}$ stacks, these will simply be more self-attractive stacks than their abelian counterpart.  

Coming back to the flux vacua respecting \eqref{su2cond}, it is important to note that the tension of the non-abelian stack gets lowered by an amount much smaller than its abelian value: indeed, the second term in the RHS of \eqref{tausu2} is a second order term in the perturbative expansions. In that sense it is a small decay channel, triggering the decay of flux vacua with abelian domain-walls that are extremal or very close to being extremal.

\subparagraph{Consequences for supersymmetric vacua} Considering these non-abelian $\mathfrak{su}(2)\oplus\mathbb{R}^{7-d}$ stacks in the case of supersymmetric vacua leads to the following statement.
\be\text{{\textit{AdS$_d$ supersymmetric vacua can't have both an $H$-flux satisfying $\sum_{i=1}^3\sqrt{\frac{s_i}{h^2}}\leq1$}}}\nonumber\ee
\be\text{{\textit{and stable BPS abelian D$(d-2)$ branes.}}}\tag{SUSY}\label{SUSYstat}\ee
Indeed, if an $H$-flux satisfying \eqref{su2cond} and stable D$(d-2)$ BPS abelian branes coexisted,\footnote{By stable abelian brane we mean here that the brane has positive second derivatives of the abelian potential.} the vacua would be destabilised by their non-abelian $\mathfrak{su}(2)\oplus\mathbb{R}^{7-d}$ counterpart and hence wouldn't be supersymmetric in full string theory, it would only be as a supergravity solution. In the case of vacua with constant warp factor and trivial dilaton profile, the condition $\sum_{i=1}^3\sqrt{\frac{s_i}{h^2}}\leq1$ is always satisfied, so $H$-flux and D$(d-2)$ BPS abelian branes can't coexist. 

To our knowledge, this behaviour is satisfied by every supersymmetric AdS flux vacuum in the literature, since $H$-flux and D$(d-2)$ BPS abelian branes never coexist in known AdS solutions.

\subsubsection{\texorpdfstring{The $\mathfrak{su}(2)\oplus\mathfrak{su}(2)$ branes}{The su(2)+su(2) branes}}

We focus here on $d\leq4$, and consider the non-abelian stack of branes obeying an $\mathfrak{su}(2)\oplus\mathfrak{su}(2)\oplus\mathbb{R}^{4-d}$ algebra, so keeping six non-vanishing scalars.
As before, we divide them into two independent families living on three-dimensional subspaces $[\Phi^\alpha,\Phi^A]=0$, with the transverse indices splitting as $i=A,B,C,\alpha,\beta,\gamma$. On each three-dimensional subspace, we have
\begin{subequations}
\begin{alignat}{2}
    f_1&=0\qquad &&h_1\epsilon_{ABC}=H_{ABC}\\
   f_2&=0\qquad  &&h_2\epsilon_{\alpha\beta\gamma}=H_{\alpha\beta\gamma}.
\end{alignat}
\end{subequations}
We consider the case where the vacua satisfy
\begin{align}S_{A\alpha}=H_{A\alpha\beta}=H_{\alpha AB}=0,\label{triveomint}
\end{align}
which is the domain-wall with internal fuzziness version of the conditions \eqref{triveom}. Such a stack of branes with scalars obeying an $\mathfrak{su}(2)\oplus\mathfrak{su}(2)\oplus\mathbb{R}^{4-d}$ algebra would therefore be a solution of the equations of motion if the NS-field-strength of the vacua has $H=H_1+H_2$, with $H_i$ proportional to the volume form of a three-dimensional subspace. This is a very restrictive requirement, to which we will come back to when we illustrate our constructions by embedding them into concrete vacua examples.

For $t_1\geq2$, $t_2\geq2$ and $s_i\geq0$, the equations of motion are therefore solved with the scalars satisfying the $\mathfrak{su}(2)\oplus\mathfrak{su}(2)$ algebra 
\begin{subequations}\label{isosu2su2}
\begin{align} [\Phi^\alpha,\Phi^A]&=0\\
[\Phi^A,\Phi^B]&=-\frac{ih_1}{2}{M_1^A}_P{M_1^B}_Q{\epsilon^{PQ}}_R{(M_1^{-1})^R}_C\Phi^C\\
[\Phi^\alpha,\Phi^\beta]&=-\frac{ih_2}{2}{M_2^\alpha}_\delta{M_2^\beta}_\mu{\epsilon^{\delta\mu}}_\nu{(M_2^{-1})^\nu}_\gamma\Phi^\gamma,
\end{align}
\end{subequations}
if and only if
\be \sum_{i=1}^3\sqrt{\frac{s_{1i}}{h_1^2}}\leq1,\qquad\sum_{i=1}^3\sqrt{\frac{s_{2i}}{h_2^2}}\leq1\label{su2su2cond}.\ee
 Again using the usual spin-$j$ $\mathfrak{su}(2)$ irrep, the effective action for the branes respecting \eqref{isosu2su2} in vacua satisfying \eqref{triveomint} is
 \begin{align} S=& L^{d-1}T_{d-2} {\rm Vol}(S^{d-1})\Big[-Ne^{(d-1)A-\phi}\sinh^{d-1} r+Lc(r) e^{dA} N(\ast\sigma F)_0\nonumber\\
 &+\frac{\lambda^2}{3\cdot 2^7}N(N^2-1)e^{(d-1)A-\phi}\sinh^{d-1} r\left(\frac{\mathrm{det}P_1}{\mathrm{det}g_1}h_1^4(t_1-2)+\frac{\mathrm{det}P_2}{\mathrm{det}g_2}h_2^4(t_2-2)\right)\Big],\label{actionsu2su2}
\end{align}
with $\mathrm{det}g_{1,2}$ the pull-back on the corresponding three-dimensional subspace. This action simply contains the two $\mathfrak{su}(2)$ contributions.
The charge and tension of the non-abelian domain walls are again read off \eqref{eq:s-inst}, they are
\begin{align}
    qf&=T_{d-2}e^{dA} L N(\ast\sigma F)_0\\
    \tau&=T_{d-2}e^{(d-1)A-\phi} N\Big[1-\frac{\lambda^2}{3\cdot 2^7}(N^2-1)\left(\frac{\mathrm{det}P_1}{\mathrm{det}g_1}h_1^4(t_1-2)+\frac{\mathrm{det}P_2}{\mathrm{det}g_2}h_2^4(t_2-2)\right)\Big],\label{tausu2su2}
\end{align}
We witness the same behaviour as before: the charge is unchanged with respect to the abelian case, while the tension gets modified. Since det$P_{1,2}>0$, when the conditions \eqref{su2su2cond} are satisfied, the stack is rendered less self-attractive than its abelian counterpart. If both fail, the possible solutions to the equations of motion have both $t_1<2$ and $t_2<2$, and the brane is more self-attractive than in the abelian case. If only one of these conditions is met, the situation gets decided case by case by the competition of both contributions.

\subsubsection{\texorpdfstring{The $\mathfrak{su}(3)$ branes}{The su(3) branes}}

We now focus on AdS$_2$ vacua and briefly consider the possibility of $\mathfrak{su}(3)$ branes, with eight non-vanishing scalars spanning the whole internal space. 

For vacua with constant dilaton and warp factor, the solution to the equations of motion \eqref{eomint} is
\be [\Phi^i,\Phi^j]=-\frac{i}{2}{H^{ij}}_k\Phi^k.\ee
As we discussed in the previous sections, such vacua always admit $\mathfrak{su}(2)$ branes. $\mathfrak{su}(2)\oplus\mathfrak{su}(2)$ branes only exists for vacua with an NS field-strength that is the sum of two three-dimensional components with independent support. Here the requirement on the NS field-strength is even more strict, since a basis in which the $H$-flux is proportional to the $\mathfrak{su}(3)$ structure constants must exist. As it is already challenging to realise AdS$_2\times M_8$ vacua with an NS field-strength having support on the whole of $M_8$, we find the possibility of $\mathfrak{su}(3)$ branes very unlikely, and don't pursue further the option of non-trivial dilaton and warp factor.
\subsubsection{\texorpdfstring{Beyond $D(d-2)$ branes}{Beyond D(d-2) branes}}
The non-abelian brane configurations discussed in  this section can be straightforwardly generalised to domain-wall branes wrapping a non-trivial internal cycle $\Sigma$ of dimension $p$: the conditions \eqref{gencond} are simply stronger on the internal geometry. Such domain-walls exist at least for vacua of the type AdS$_d\times\Sigma\times M_{10-d-p}$, with $M$ any manifold.
\subsection{Radial fuzziness}
We now relax the condition of purely internal fuzziness and consider both internal scalars and a non-vanishing scalar along the radial direction of AdS$_d$. We are still within the framework of the previous section, that is we respect the conditions \eqref{gencond}, and we keep on focusing on D$(d-2)$ domain-wall branes, with \eqref{CRR} as our choice of RR potential.

The compactification ansatz and the isometries of AdS still ensure $E_{ai}=0$ and render the choice of trivial gauge $A_a=0$ natural, so the only restriction on our domain-wall stack is that it is homogeneous in the transverse directions $\partial_a\Phi^i=0$.

However, a non-vanishing radial scalar has drastic consequences: the stack of branes must now sit at a stationary point of the abelian potential in the radial direction. The corresponding condition \eqref{min} is
\be\tanh{r}=\frac{(d-1)e^{-A-\phi}}{L(\ast\sigma F)_0}.\label{minrad}\ee
The tension and charge of an abelian D$(d-2)$ domain-wall brane are $\tau=T_{d-2}e^{(d-1)A-\phi}$ and $qf=LT_{d-2}e^{d A}(\ast\sigma F)_0$, respectively \cite[Sec.~2]{Menet:2025nbf}. This requirement \eqref{minrad} is therefore nothing but the condition for the existence of an Euclidean instanton \eqref{instrad}.\footnote{In the case of AdS$_3$ with an external $H$-flux, the radius of the instanton is instead dictated by \eqref{instrB}.} This means that requiring the non-abelian stack to sit at the position of an abelian extremum can only be achieved if the abelian domain-wall destabilises the vacua it sits in. If one thinks in terms of vacuum decay, this makes the stack of branes with both radial and internal fuzziness less interesting than the purely internally fuzzy one, as it destabilises vacua that could also be destabilised by their abelian counterpart. 

    However, as we will see shortly, it turns out that these non-abelian configurations can exist, and be energetically favored over their abelian homologues. They are therefore the fastest decay channel that destabilizes the vacua. Those that saturate the stability bound have an interpretation as non-abelian vacua of the boundary CFT. The fact that their fuzziness involves the radial coordinate means that the energy scale is also involved. The exotic nature of these non-abelian branes is worth investigating further.
    
\subsubsection{\texorpdfstring{The $\mathfrak{su}(2)$ branes}{The su(2) branes}}
We focus on the $\mathfrak{su}(2)\oplus\mathbb{R}^{7-d}$ possibility, and keep only three non-vanishing scalars. Two are internal, and the last one is the radial scalar.
Once again we use the potential \eqref{CRR}, and the RR equations of motion \eqref{RReq} simplify the three-dimensional top-forms \eqref{3d} to 
\be f=e^{dA+\phi}L\, \iota_1\iota_2(\ast\sigma F)|,\qquad h=0,\ee
where we set $\epsilon_{r12}=1$. \eqref{sij} and \eqref{minrad} yield\footnote{In the case of AdS$_3$ with an external $H$-flux $h$, $S_{rr}$ is a cumbersome function of $r$ and $h$, which we don't display here.}
\begin{align} S_{rr}&=d - 1\,,
\label{Srrf0}
\end{align}
while \eqref{sij} and \eqref{RReq} give 
\be 
S_{ij}=\frac{2}{d-1}\partial_i\partial_j\varphi\, \quad i,j=1,2,\qquad S_{r1}=0\qquad S_{r2}=0.
\ee
We see here that $S_{rr}$ is always positive. The radial direction is therefore never a tachyonic direction of the abelian potential.

    Let us consider solutions with $t>2$. That is, non-abelian $\mathfrak{su}(2)$ stacks that are energetically favored over their abelian counterpart. In the case of constant warp factor and dilaton, the condition \eqref{su2condgen} simplifies to
\label{boundrad}
\begin{align}
 d-1&\leq f^2\label{boundposrad}.
 \end{align}
 We will come back to this condition in the next section, when we embed these radial $\mathfrak{su}(2)$ branes into concrete flux vacua.

\subsubsection{\texorpdfstring{The $\mathfrak{su}(2)\oplus\mathfrak{su}(2)$ branes}{The su(2)+su(2) branes}}
We consider again the case of a non-abelian stack of branes obeying an $\mathfrak{su}(2)\oplus\mathfrak{su}(2)\oplus\mathbb{R}^{4-d}$ algebra, so keeping six non-vanishing scalars. One triplet of scalars is the one we just discussed, mixing the radial and two internal directions, while the other is a fully internal one. The inequalities to respect to admit each energetically favored $\mathfrak{su}(2)$ stack are as above, while the complete $\mathfrak{su}(2)\oplus\mathfrak{su}(2)\oplus\mathbb{R}^{4-d}$ algebra is obeyed by the six scalars if the additional equations of motion \eqref{su2su2eom} are satisfied.
We only consider the case where we have 
\begin{subequations}\label{triveom2}
\begin{align}&S_{A\alpha}=\iota_{[A}\iota_\alpha\iota_{\beta]}F_{10}|=H_{A\alpha\beta}=0\\
&S_{\alpha A}=\iota_{[\alpha}\iota_A\iota_{B]}F_{10}|=H_{\alpha AB}=0.
\end{align}
The additional $\mathfrak{su}(2)\oplus\mathfrak{su}(2)\oplus\mathbb{R}^{4-d}$ equations of motion are then trivially satisfied.
\end{subequations}
\section{Flux vacua examples}
We illustrate the results of the previous sections by embedding our non-abelian branes into some concrete flux vacua.
\subsection{\texorpdfstring{AdS$_4\times \mathbb{CP}^3$ and AdS$_4\times \mathbb{F}(1,2;3)$ vacua}{AdS4xCP3 and AdS4xF(1,2;3) vacua}}
We discuss the vacuum decay of some type IIA AdS$_4\times \mathbb{CP}^3$ and AdS$_4\times \mathbb{F}(1,2;3)$ vacua mediated by $\mathfrak{su}(2)$ stacks of internally fuzzy D2 domain-wall branes. 
\subsubsection{Non-supersymmetric vacua}
We consider the $\mathbb{CP}^3$ solutions first discussed in \cite{Koerber:2010rn}, and we follow the construction and conventions of \cite[Sec.~4.1]{Menet:2025nbf}. These are families of solutions defined by a shape parameter $\sigma$.

\begin{figure}[!ht]
   \centering
\includegraphics[width=0.5\linewidth]{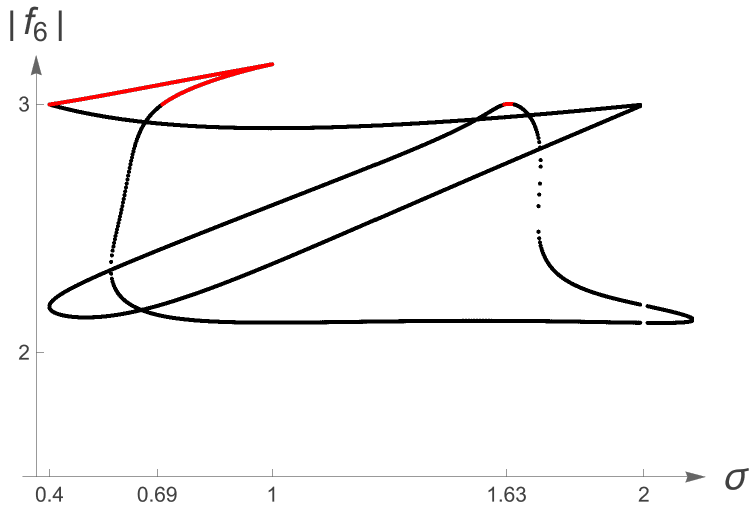} 
   \caption{\small The $f_6$ flux for various $\mathbb{CP}^3$ vacua. The red branches correspond to the vacua destabilised by abelian domain-wall D2 branes. The solutions in black are stable against the nucleation of these branes.}
    \label{D2fig}
\end{figure}

Let us mention the case of an abelian D2 domain-wall brane first. In this language, the bound \eqref{eq:dq-stab} for the vacua to be destabilised by such a D2 brane is\footnote{$L$ and $g_s$ have been absorbed into the definition of the RR flux here.}
\be |f_6|>3,\label{boundcp3}\ee
with $f_6$ the RR flux along the volume-form of AdS$_4$.
Several non-supersymmetric solutions have been found numerically in \cite{Koerber:2010rn}, and the values of $f_6$ for these solutions are displayed in figure \ref{D2fig}, adapted from \cite{Menet:2025nbf}.

    We consider now an $\mathfrak{su}(2)$ stack of internally fuzzy D2 domain-wall branes, with three non-vanishing internal scalars. These can be any three out of the six internal dimensions. However, the NS field-strength of these solutions is $H=h\mathrm{Re}\Omega$, with $\Omega$ defined by the internal SU$(3)$ structure. We therefore pick three directions such that the pull-back of $H$ to the three-dimensional subspace they define doesn't vanish.

    These solutions have no warping and a trivial dilaton profile, they therefore admit an energetically favored internally fuzzy $\mathfrak{su}(2)$ domain-wall D2 stack with $P=\mathbb{I}$ and $t=3$, as discussed in section \ref{su2intern}. From the charge and tension of the stack \eqref{qtausu2}, the instability bound \eqref{eq:dq-stab} becomes
    \be |f_6|>3\Big(1-\frac{\lambda^2h^4}{3\cdot 2^7}(N^2-1)\Big).\ee
    The second term on the right-hand side is negative, this decay channel therefore requires an $|f_6|$ smaller than its abelian counterpart. 

    Looking back at figure \ref{D2fig}, we conclude that the solutions that are stable against the abelian D2 decay channel, in the neighbourhoods of $\sigma=0.69$ and $\sigma=1.63$, are now destabilised by this non-abelian stack. 
    
    However, this destabilisation only occurs in the vicinity of $|f_6|=3$, since we have that $\frac{\lambda^2h^4}{3\cdot 2^7}(N^2-1)\ll1$. In that sense, this is a small decay channel. It is worth mentioning nonetheless that the 
    $\mathbb{CP}^3$ vacua that are destabilised by this channel were previously resisting all abelian branes and bound-states decay \cite{Menet:2025nbf}.

    The exact same situation occurs for the $\mathbb{F}(1,2;3)$ vacua of \cite{Menet:2025nbf}.
\subsubsection{Supersymmetric vacua}
Let us quickly illustrate how the statement \eqref{SUSYstat}, made in section \ref{su2intern}, plays out in the case of the supersymmetric $\mathbb{CP}^3$ solutions. They are a family of solutions on $\sigma\in[\frac25,2]$, and they satisfy
\be |f_6|=\frac{3(2+\sigma)}{4},\qquad H=-\frac{1}{2}\sqrt{(5\sigma-2)(2-\sigma)}.\ee
The abelian D2 brane is not BPS for generic values of $\sigma$, and $H$ is non-vanishing in general. However, for $\sigma=2$, $|f_6|=3$ is BPS, and the $H$-flux vanish, illustrating that the existence of BPS D2 domain-walls and $H$-flux is mutually conflicting for supersymmetric AdS$_4$ vacua.
\subsection{Radially fuzzy branes in symmetric vacua}
We discuss the embedding of radial $\mathfrak{su}(2)\oplus\mathbb{R}^{4}$ stacks into the symmetric solutions \cite{Wulff:2017zbl} of type II supergravity.

To illustrate, let us focus on the type IIA AdS$_4\times S^4\times S^2$ vacua with the following fluxes\cite[(2.16)]{Wulff:2017zbl}:
\begin{alignat}{2}
H&=0,\qquad \quad F^{10}&&=f_4+f_2\mathrm{Vol}_{S^4}+f_3\mathrm{Vol}_{S^2}-f_1\mathrm{Vol}_{S^4}\wedge\mathrm{Vol}_{S^2}\nonumber\\
&\; &&\quad+\mathrm{Vol}_{\mathrm{AdS}_4}\wedge\big(f_1+f_2\mathrm{Vol}_{S^2}-f_3\mathrm{Vol}_{S^4}+f_4\mathrm{Vol}_{S^4}\wedge\mathrm{Vol}_{S^2}\big)\\
f_2&=\frac{f
_3f_4}{f_1},\qquad f_1^2&&=\frac{1}{2} \left( 5 f_4^2 + 3 f_3^2 + \sqrt{\left(5 f_4^2 + 3 f_3^2\right)^2 + 4 f_4^2 f_3^2} \right).
\end{alignat}
This vacuum has no warping and a trivial dilaton profile, which has been absorbed into the RR fields, together with the AdS length scale. Similarly, we parametrise the curvature of the two-sphere as $R_{mn}\equiv\lambda\, g_{mn}$, and we have
\be f_3 = \sqrt{\frac{3}{2\lambda}}\sqrt{-6+\lambda+\lambda^2}\qquad f_4=\frac{1}{\sqrt{2\lambda}}\sqrt{18-3\lambda-\lambda^2}.\ee
The reality of the fluxes imposes $\lambda\in[2,3]$, for which we have $f_1\geq3$. There is therefore a destabilizing abelian D2 brane, which is a necessary condition to admit radially fuzzy non-abelian branes.  
We consider non-vanishing scalars along the two-sphere and the radial direction of AdS$_4$. The only non-zero entry of the matrix $S$ is the radial component. The equation of motion for the scalars, here in the form of the bound \eqref{boundposrad}, then becomes
\be \frac{3}{2\lambda}(3-\lambda)(2-\lambda)\geq3\label{boundradW}.\ee
This is never satisfied for $\lambda\in[2,3]$, this vacuum thus never admits radially fuzzy D2 branes.

It turns out that this behavior is common to all symmetric solutions of \cite{Wulff:2017zbl}: when the fluxes are such that the abelian branes are super-extremal, the equations of motion for the radial scalar is never satisfied, and the vacua never admit radially fuzzy branes.


\subsection{\texorpdfstring{AdS$_3\times S^3\times S^3\times S^1$ vacua}{AdS3xS3xS3xS1 vacua}}
We illustrate the construction of $\mathfrak{su}(2)\oplus\mathfrak{su}(2)$ stacks of domain-wall branes with purely internal fuzziness in AdS$_3\times S^3\times S^3\times S^1$ vacua.
\subsubsection{Type IIB}
We first focus on $\mathfrak{su}(2)\oplus\mathfrak{su}(2)$ stacks of D1 domain-walls with purely internal fuzziness, embedded in the AdS$_3\times S^3\times S^3\times S^1$ vacua with fluxes \cite[(2.80)]{Wulff:2017zbl}
\begin{align}
H&=f_1(\mathrm{Vol}_{\mathrm{AdS}_3}+f_2\mathrm{Vol}_{S^3_1}+f_3\mathrm{Vol}_{S^3_2}),\label{Hsu2su2}\\
F^{10}&=f_4\Big(f_2\mathrm{Vol}_{S^3_1}+f_3\mathrm{Vol}_{S^3_2}-\mathrm{Vol}_{S^3_1}\wedge\mathrm{Vol}_{S^3_2}\wedge\mathrm{Vol}_{S^1}\nonumber\\
&\quad+\mathrm{Vol}_{\mathrm{AdS}_3}\wedge(1-f_2\mathrm{Vol}_{S^3}\wedge\mathrm{Vol}_{S^1}-f_3\mathrm{Vol}_{S^3}\wedge\mathrm{Vol}_{S^1})\Big),
\end{align}
with $f_2^2+f_3^2=1$.

 We consider the six non-vanishing scalars to be along the two three-spheres. Again, this vacuum has no warping and a trivial dilaton profile, it therefore admits $\mathfrak{su}(2)$ domain-wall D1 stacks with internal fuzziness along each three-sphere. The additional equations of motion to admit an $\mathfrak{su}(2)\oplus\mathfrak{su}(2)$ stack are here   
\begin{align}H_{A\alpha\beta}=H_{\alpha AB}=0,\label{triveomintads3}
\end{align}
with the Latin indices for one three-sphere and the Greek ones for the other. These are satisfied by the $H$-flux \eqref{Hsu2su2}, so this vacuum admits an $\mathfrak{su}(2)\oplus\mathfrak{su}(2)$ stack of D1 domain-wall branes, that are fuzzy along the two three-sphere directions. 

The D1 abelian domain-wall will be unstable if the bound \eqref{instabB} is satisfied. Taking the external Einstein equation $f_1^2+f_4^2=4$ into account, this bound here takes the form $|f_4|>\sqrt{f_4^2+2f_1^2}$, which cannot be satisfied. These vacua are therefore always stable against the nucleation of abelian D1 domain-walls. The extreme case $f_1=0$ leaves the abelian D1 extremal, but then no non-abelian contributions are available since $H=0$. When $f_1\neq0$, non-abelian D1 stacks could however still nucleate and expand in principle, since the instability bound becomes 
\be|f_4|>\sqrt{f_4^2+2f_1^2/4}\left(1-\lambda^2f_1^4\frac{f_2^4+f_3^4}{3\cdot 2^7}(N^2-1) \right).\label{boundS3S3S1}
\ee
For example, if one considers vacua such that $f_4^2\approx3.99,\, f_1^2\approx0.01,\, f_2^4+f_3^4\approx 1$ and $l_s\approx10^{-2}$, the instability bound is now $N\gtrsim 10^5$. A value of $N$ satisfying this bound reasonably is such that the regime of validity for the $\mathfrak{su}(2)\oplus\mathfrak{su}(2)$ stack \eqref{regimesu2su2} is satisfied, and it destabilises the vacua where the abelian D1 failed to do so.

    It is interesting to note that an $\mathfrak{su}(2)$ brane with fuzziness along only a single sphere could also exist and destabilise the vacua, but only $f_2$ or $f_3$ would be entering the analogue of \eqref{boundS3S3S1}, depending on which sphere is fuzzy. This brane stack would therefore be more self-attractive than the $\mathfrak{su}(2)\oplus\mathfrak{su}(2)$ one, and would thus be slower at triggering the vacuum decay.
\subsubsection{Type IIA}
We consider again the construction of an $\mathfrak{su}(2)\oplus\mathfrak{su}(2)$ stack of domain-wall branes with purely internal fuzziness in AdS$_3\times S^3\times S^3\times S^1$ vacua. However, this time the stack wraps a non-trivial internal cycle, illustrating how this construction goes beyond purely domain-wall branes.

We consider the T-dual of the previous vacua \cite[(2.33)]{Wulff:2017zbl} 
\begin{align}
H&=f_1(\mathrm{Vol}_{\mathrm{AdS}_3}+f_2\mathrm{Vol}_{S^3_1}+f_3\mathrm{Vol}_{S^3_2}),\label{Hsu2su2IIA}\\
F^{10}&=f_4\Big(f_2\mathrm{Vol}_{S^3_1}\wedge\mathrm{Vol}_{S^1}+f_3\mathrm{Vol}_{S^3_2}\wedge\mathrm{Vol}_{S^1}-\mathrm{Vol}_{S^3_1}\wedge\mathrm{Vol}_{S^3_2}\nonumber\\
&\quad+\mathrm{Vol}_{\mathrm{AdS}_3}\wedge(\mathrm{Vol}_{S^1}+f_3\mathrm{Vol}_{S^3_1}+f_2\mathrm{Vol}_{S^3_2})\Big),
\end{align}
with $f_2^2+f_3^2=1$. We now examine domain-wall D2-branes wrapping the internal $S^1$. The six non-vanishing scalars are the remaining internal directions, the two three-spheres. As before, the absence of warping and the trivial dilaton profile ensure that the vacuum admits $\mathfrak{su}(2)$ D2 domain-walls with internal fuzziness along each three-spheres. The remaining equations of motion for the $\mathfrak{su}(2)\oplus\mathfrak{su}(2)$ stack are again   
\begin{align}H_{A\alpha\beta}=H_{\alpha AB}=0,
\end{align}
and are satisfied by the $H$-flux \eqref{Hsu2su2IIA}. This vacuum therefore admits an $\mathfrak{su}(2)\oplus\mathfrak{su}(2)$ stack of D2 domain-wall branes wrapping $S^1$, which are fuzzy along the two three-sphere directions. 
\section{Discussion}

In this paper, we studied non-abelian D$p$-branes in curved space, using the Myers action. These branes are weakly non-abelian, in the sense that we only considered contributions to the action of order up to $\alpha'^2$. We first wrote down the equations of motion for the matrix-valued transverse scalars in the case where they were constant along the world-volume directions. We then solved these equations for the cases where the constant scalars obey an $\mathfrak{su}(2)\oplus\mathbb{R}^{6-p}$ algebra. By this we mean that we gave criteria on the background fluxes that the vacua must respect to admit such branes. We also derived some conditions on flux vacua to admit non-abelian branes satisfying an $\mathfrak{su}(2)\oplus\mathfrak{su}(2)\oplus\mathbb{R}^{6-p}$ algebra. We focused on the non-abelian brane configurations that were energetically favored over their abelian counterpart.

We then specialised our results to the case of non-abelian D$(d-2)$ domain-wall branes in AdS$_d$ vacua, motivated by the consequences these branes could have on the stability of non-supersymmetric AdS vacua. We found that the $\mathfrak{su}(2)\oplus\mathbb{R}^{6-p}$ and $\mathfrak{su}(2)\oplus\mathfrak{su}(2)\oplus\mathbb{R}^{6-p}$ solutions we considered have the same charge than their abelian cousins, but a lower tension, rendering them less self-attractive. This has interesting implications for the vacua that resist the abelian domain-wall decay channels, since they can now be destabilised by these branes. However, since we treat only weakly non-abelian branes, this new decay channel requires the abelian branes homologues to be close to extremality or extremal.

We have constructed two types of non-abelian domain-wall branes: the ones with a radial fuzziness, along the radial directions of AdS, and the ones without. The branes with radial fuzziness must be expanded around superextremal abelian branes, while the purely internally fuzzy branes are free from this constraint. The radially fuzzy branes therefore don't entail a \emph{new} decay channel, per se.

We applied our construction to some non-supersymmetric AdS vacua that were resisting all their assessed abelian decay-channels, and destabilised some of them, most notably some AdS$_4\times \mathbb{CP}^3$ and AdS$_4\times \mathbb{F}(1,2;3)$ vacua which have abelian D2 branes close to extremality.

It is worth discussing the implications of this construction for supersymmetric vacua: if they have extremal abelian branes, they can't develop the non-abelian branes we discussed. This makes our criterion on the background fluxes \eqref{su2condgen} and the presence of extremal abelian D$p$-branes mutually conflicting. The clearest example for illustration is the one of vacua without warping and with a trivial dilaton profile, together with a stable BPS abelian D$p$-brane. In this case our condition \eqref{su2condgen} is automatically satisfied. The mere coexistence of an $H$-flux and such a brane is then forbidden. The stability of supersymmetric vacua is therefore here rephrased as constraints on the collective behaviour of branes and fluxes. Alternatively, if these constraints are violated by a supersymmetric solution of supergravity, it could indicate that this vacuum isn't a truly supersymmetric solution of string theory, as the non-abelian contributions to the branes tension are stringy corrections.

One obvious extension of this work is to further study the constraints \eqref{su2condgen} and \eqref{su2condgenneg} for vacua with warping and non-trivial dilaton. In this direction, the DGKT-like AdS$_4$ vacua dubbed \textbf{A1-S1} in \cite{Marchesano:2019hfb} are particularly appealing subjects. They have a supersymmetric and a non-supersymmetric branch, and both hold an extremal abelian D$4$ domain-wall, which remains extremal after localisation  \cite{Marchesano:2022rpr} (to low order in $g_s$). It would be interesting to see if our constraints to admit $\mathfrak{su}(2)$ branes hold here such that the non-supersymmetric vacuum is destabilised by non-abelian D$4$-branes. \emph{More speculatively}, if the supersymmetric branch also satisfies these constraints, it could signal that this DGKT-like vacuum isn't a genuine supersymmetric vacuum of string theory, as discussed above. The possibility that supersymmetric DGKT-like vacuum could actually be non-supersymmetric in full string theory has been discussed before in \cite{Montero:2024qtz}. 
\section*{Acknowledgments}
We are supported in part by the INFN, and by the MUR-PRIN contract 2022YZ5BA2.

\bibliographystyle{JHEP}
\bibliography{biblio}

\end{document}